%% file: bare_conf.tex
\documentclass[conference]{IEEEtran}
\usepackage{cite}
\ifCLASSINFOpdf
\else
\fi
\usepackage{algorithmic}
\usepackage[]{algorithm}

\usepackage{tikz}

\input{header}

\begin{document}
%
% paper title
% Titles are generally capitalized except for words such as a, an, and, as,
% at, but, by, for, in, nor, of, on, or, the, to and up, which are usually
% not capitalized unless they are the first or last word of the title.
% Linebreaks \\ can be used within to get better formatting as desired.
% Do not put math or special symbols in the title.
%        10        20        30        40        50        60        70        80        90       100       110
%123456789012345678901234567890123456789012345678901234567890123456789012345678901234567890123456789012345678901
%Base-Stations Up in the Air: Multi-UAV Trajectory Control for Max-Min-Rate Optimization in Uplink C-RAN Systems (111)
%Base-Stations Up in the Air: Multi-UAV Trajectory Control for Min-Rate Maximization in Uplink C-RAN Systems (107)
%Flying Base-Stations: Multi-UAV Trajectory Control for Max-Min-Rate Optimization in Uplink C-RAN Systems (104)
%Flying Base-Stations: Multi-UAV Trajectory Control for Min-Rate Maximization in Uplink C-RAN Systems (100)
%Base-Stations Up in the Air: Multi-UAV Trajectory Control for Min-Rate Maximization in Uplink C-RAN (99)
\title{Base-Stations Up in the Air:\\Multi-UAV Trajectory Control for Min-Rate Maximization in Uplink C-RAN}

%\title{Flying Base-Stations:\\Multi-UAV Trajectory Control for Min-Rate Maximization in Uplink C-RAN Systems}

%\title{Base-Stations Up in the Air:\\ Multi-UAV Trajectory Control for Max-Min-Rate Optimization in Uplink C-RAN Systems}

% author names and affiliations
% use a multiple column layout for up to three different
% affiliations
%\author{%
%\IEEEauthorblockN{Stefan Roth}
%\IEEEauthorblockA{Digital Communication Systems\\
%Ruhr University Bochum\\
%Email: stefan.roth-k21@rub.de}
%\and
%\IEEEauthorblockN{Ali Kariminezhad}
%\IEEEauthorblockA{Digital Communication Systems\\
%	Ruhr University Bochum\\
%	Email: ali.kariminezhad@rub.de}
%\and
%\IEEEauthorblockN{Aydin Sezgin}
%\IEEEauthorblockA{Digital Communication Systems\\
%	Ruhr University Bochum\\
%	Email: aydin.sezgin@rub.de}
%}
\author{%
\IEEEauthorblockN{Stefan Roth, Ali Kariminezhad, and Aydin Sezgin}
\IEEEauthorblockA{Institute of Digital Communication Systems, 
	Ruhr University Bochum, Germany\\
	Email: \{stefan.roth-k21,ali.kariminezhad,aydin.sezgin\}@rub.de}
}

% conference papers do not typically use \thanks and this command
% is locked out in conference mode. If really needed, such as for
% the acknowledgment of grants, issue a \IEEEoverridecommandlockouts
% after \documentclass

% for over three affiliations, or if they all won't fit within the width
% of the page, use this alternative format:
% 
%\author{\IEEEauthorblockN{Michael Shell\IEEEauthorrefmark{1},
%Homer Simpson\IEEEauthorrefmark{2},
%James Kirk\IEEEauthorrefmark{3}, 
%Montgomery Scott\IEEEauthorrefmark{3} and
%Eldon Tyrell\IEEEauthorrefmark{4}}
%\IEEEauthorblockA{\IEEEauthorrefmark{1}School of Electrical and Computer Engineering\\
%Georgia Institute of Technology,
%Atlanta, Georgia 30332--0250\\ Email: see http://www.michaelshell.org/contact.html}
%\IEEEauthorblockA{\IEEEauthorrefmark{2}Twentieth Century Fox, Springfield, USA\\
%Email: homer@thesimpsons.com}
%\IEEEauthorblockA{\IEEEauthorrefmark{3}Starfleet Academy, San Francisco, California 96678-2391\\
%Telephone: (800) 555--1212, Fax: (888) 555--1212}
%\IEEEauthorblockA{\IEEEauthorrefmark{4}Tyrell Inc., 123 Replicant Street, Los Angeles, California 90210--4321}}

% use for special paper notices
%\IEEEspecialpapernotice{(Invited Paper)}

% make the title area
\maketitle

% As a general rule, do not put math, special symbols or citations
% in the abstract
\begin{abstract}
In this paper we study the impact of unmanned aerial vehicles (UAVs) trajectories on terrestrial users' spectral efficiency (SE). Assuming a strong line of sight path to the users, the distance from all users to all UAVs influence the outcome of an online trajectory optimization. The trajectory should be designed in a way that the fairness rate is maximized over time. That means, the UAVs travel in the directions that maximize the minimum of the users' SE. From the free-space path-loss channel model, a data-rate gradient is calculated and used to direct the UAVs in a long-term perspective towards the local optimal solution on the two-dimensional spatial grid. Therefore, a control system implementation is designed. Thereby, the UAVs follow the data-rate gradient direction while having a more smooth trajectory compared with a gradient method. The system can react to changes of the user locations online; this system design captures the interaction between multiple UAV trajectories by joint processing at the central unit, e.g., a ground base station. Because of the wide spread of user locations, the UAVs end up in optimal locations widely apart from each other. Besides, the SE expectancy is enhancing continuously while moving along this trajectory.
\end{abstract}
% no keywords

% For peer review papers, you can put extra information on the cover
% page as needed:
% \ifCLASSOPTIONpeerreview
% \begin{center} \bfseries EDICS Category: 3-BBND \end{center}
% \fi
%
% For peerreview papers, this IEEEtran command inserts a page break and
% creates the second title. It will be ignored for other modes.
\IEEEpeerreviewmaketitle

\section{Introduction}
%This demo file is intended to serve as a ``starter file''
%for IEEE conference papers produced under \LaTeX\ using
%IEEEtran.cls version 1.8b and later.
% You must have at least 2 lines in the paragraph with the drop letter
% (should never be an issue)
% no \IEEEPARstart
% TODO Profe cite 7832322 for useful things.
% UAV-aided ubiquitous coverage
%
% As I understand, the drop letter is only relevant for journal papers. Thus, not applicable. \IEEEPARstart{F}{uture}
Future communication systems is expected to be responsive to a plethora of users due to integrating new concepts such as internet of things (IoT) and large-scale sensor networks. Furthermore, the quality of service (QoS) demands of the users is a continuously increasing trend. For improving the QoS, drones also known as unmanned aerial vehicles (UAVs), are proposed to be deployed in communication networks. The authors in~\cite{8316776} study the integration of UAVs into 5G and beyond 5G (B5G) cellular networks as they enable the possibility of an additional line of sight (LoS) connection. %Involving UAVs in communication networks requires addressing some challenges.

%DBLP:journals/corr/abs-1803-00680: A Tutorial on UAVs for Wireless Networks: Applications, Challenges, and Open Problems
%DBLP:journals/corr/abs-1711-07668:  Massive MIMO for Drone Communications: Applications, Case Studies and Future Directions
%8292533: DBLP:journals/corr/abs-1805-07822: UAV-aided Multi-Way Communications
%Waveform and spectrum management for unmanned aerial systems beyond 2025
Involving UAVs in communication networks requires addressing some key challenges that have been identified by \cite{DBLP:journals/corr/abs-1803-00680} and \cite{DBLP:journals/corr/abs-1711-07668}. They contain air-to-ground channel modeling\cite{8048502}, resource and trajectory optimization \cite{8376956}, and spectrum management\cite{8292533}. Moreover, cellular networks need to be planned \cite{DBLP:journals/corr/abs-1805-07822} as a new MAC layer design is required \cite{5461502} and security issues need to be addressed\cite{8255739}. In this paper, trajectory and resources are optimized jointly based on the users' location.

% 6863654 altitude
% 7486987 deployment: both, horizontal and vertical
% DBLP:journals/corr/MozaffariSBD15a: Small cells, horizontal and vertical
% 7762053: horizontal
% 8376956 alernating deployment
% 8011325 FlyBS
Thereby, UAVs can move at the same time as serving terrestrial users. %This trajectory of the UAVs can then be optimized, if the location of the users are given.
%The optimal placement of UAVs horizontally and vertically are two optimization criteria for data rate maximization that can be optimized together  \cite{7486987,DBLP:journals/corr/MozaffariSBD15a} or individually.
Due to the possibility of drones for being mobile, they should be guided in directions that help the users' QoS. The authors in\cite{7486987,DBLP:journals/corr/MozaffariSBD15a} study UAV trajectory in azimuth and altitude for the aim of rate maximization. 
Interestingly, the coverage area can be controlled by UAV altitude adjustment. This adjustment captures the trade-off between an enlarged coverage area and high power consumption due to high pathloss serving the users~\cite{6863654}. The horizontal placement of a set of UAVs over the ground is optimized in \cite{7762053}. While moving, the trajectory can be obtimized for either alternating deployment towards multiple users \cite{8376956} or for maximizing constant data rates as flying base station (FlyBS) \cite{8011325}. In this paper, the latter one is considered.
Having multiple UAVs, the users' signals can either be decoded locally at the supporting UAV, or, in case of C-RAN, jointly processed at a central unit. The former requires allocating the users to UAVs, which is addressed in~\cite{8385464}. However, the latter deals with the latency challenge due to joint processing at the central unit. This challenge can be remedied by having very high capacity links in fronthaul between the UAVs and the central unit.
%Therefore, %the authors in~\cite{6863654,7486987,7762053} discuss
%the impact of UAV position on the coverage areas is discussed in~\cite{6863654,7486987,7762053} with the aim to serve stochastically located users.
%% These works are based on the stochastic models of the user locations, i.e., the UAVs are located stochastically, but the exact user locations are not taken into account.
%Moreover, the trajectory of a single UAV in two-dimensional spatial coordinates is investigated in~[]. In~[] the authors study the performance of UAV-aided communication, where the UAV is used as a relay. The optimal allocation of users to UAVs is described in \cite{8385464}.

\begin{figure}[t]
    \centering
    \includegraphics{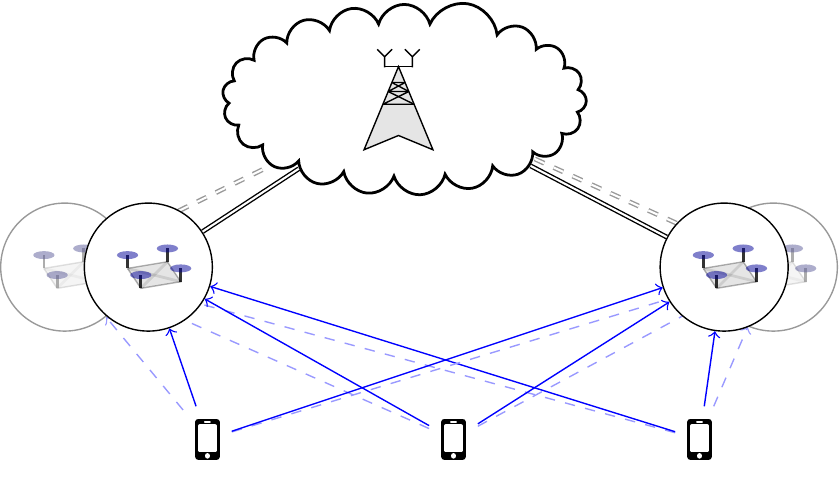}
    \caption{UAVs optimize their locations while receiving data from a set of users until an equilibrium is reached. The links towards the fronthaul network have constant high capacity.}
    \label{fig:scenario}
\end{figure}

In this paper, we consider a scenario, where multiple users demand connectivity to the ground base station (BS). Due to the large distance of the users to the ground BS, the achievable spectral efficiency (SE) is considerably low. Here, we exploit multiple UAVs forming a C-RAN, for aiding the communication between the users and the ground BS. For simplicity, we assume that the UAVs fully cooperate by considering an unlimited fronthaul capacity between the UAVs and the ground BS, e.g., optical  communication for the fronthaul links \cite{1563300}.
Due to the unlimited fronthaul capacity assumption, and given the location of the users, the optimal travel directions of all UAVs are calculated by the ground BS and then provided to all UAVs through the fronthaul links, for controlled trajectory purposes.
The optimized trajectory of a UAV is a function of the optimization utility. That means, a power-efficient trajectory differs from the trajectory which maximizes the SE. As the user locations can be highly dynamic, we investigate UAVs travelling in a way that the data rate is maximized in short time intervals in this paper.
Thus, all quadcopter UAVs travel in the direction the maximizes the SE of the users, i.e., traveling in the gradient direction. The scenario is illustrated in \figurename~\ref{fig:scenario}.

%Given the gradient information from the BS, a UAV travels based on the output of the control unit. Here, a quadcopter with a controller plays the role of a UAV.

The trajectory of controlled quadcopters is compared with a more abstract gradient method, where the gradient directly determines the flight direction and velocity, i.e., zero gradient represents zero motion. This comparison has less mathematical complexity, but is not suitable for online applications. Since the travel direction is calculated based on the user locations by the ground BS online, a change in the users' locations results in an instant change in travel direction. The optimization problem is stated in the way that the minimum rate should be maximized in section~\ref{sec:optimizationProblem}. In section~\ref{sec:numericalResults}, the performance of the control design is compared with an ideal gradient method numerically.% for solving the max-min optimization problem.

\subsection{Notation}
%\paragraph{Notation}
The following notation is used in this paper: Matrices, vectors, and sets are indicated by bold upper-case, bold lower-case, and calligraphy letters, respectively. Determinant, trace, Hermitian and transpose of a matrix $\mat{A}$ are represented by $\left|\mat{A}\right|$, $\trace\left\{\mat{A}\right\}$, $\mathbf{A}^{H}$ and $\mathbf{A}^{T}$ respectively. The cardinally of a set $\mathcal{A}$ is indicated by $\left|\mathcal{A}\right|$. The $L_2$-norm of a vector $\vect{a}$ is denoted by $\|\vect{a}\|$. The Kronecker and Hadamard products are represented by $\otimes$ and $\circ$, respectively. The function $\phase\{\mat{A}\}$ gives the element-wise complex phases of the matrix.

\section{System Model}
A set of users $\mathcal{G}$ intends to transmit data in cellular uplink to a set of UAVs $\mathcal{K}$, which are considered as flying base stations. This UAVs have a strong fronthaul connection towards a central base station on the ground and can adjust their positions while providing data rates to the users.
The number of users and UAVs are given as $G=|\mathcal{G}|$ and $K=|\mathcal{K}|$, respectively. In the following, the channel model and a quadcopter UAV system model are described.
\subsection{Channel Model}
Each user is equipped with $N_T$ antennas; each UAV has $N_R$ antennas. We consider a single dominant LoS path between the users and the UAVs.  The distance between the $k$th UAV and $i$th user is represented by
\begin{align}
    d_{\mathrm{U}k,i}=\left|\left|\vect{x}_{\mathrm{U}k}-\vect{x}_{i}\right|\right|,
\end{align}
where $\vect{x}_{\mathrm{U}k}$ and $\vect{x}_{i}$  indicate the position coordinates of the $k$th UAV, and the $i$th user, respectively.

%Due to far-field assumption, the path-loss between the transmit and receive antennas are considered identical.
Based on the free-space path-loss model, the path loss between each antenna of user $i$ and each antenna at UAV $k$ can be considered identical and is approximated by \cite{8048502}
\begin{align}
\text{PL}(d_{\mathrm{U}k,i})=10\alpha\log\left(\frac{d_{\mathrm{U}k,i}}{d_0}\right){\si{\dB}}+\text{PL}\left(d_0\right)+w_{\sigma ki}{\si{\dB}},\label{eq:pathLoss}
\end{align}
%TODO{\color{red} Possible solution for the stochastic part: remove $x_\sigma$, and describe above as expectancy of the path loss. WARNING: The cite used here references another source which is not available freely.}\\
in which, $\text{PL}(d)$ is the path-loss at distance $d$ in dB, $\alpha$ is the path loss exponent, and $d_0$ represents a reference distance. A shadowing component is represented by the Gaussian variable $w_{\sigma ki}\sim\Gaussian(0,\sigma^2)$.

Defining $\beta=\exp\left(-\frac{\ln(10)}{\SI{10}{\dB}}\text{PL}\left(d_0\right)\right)d_0^\alpha$, the channel matrix between user $i$ and UAV $k$ is approximated as
\begin{align}
\mat{H}_{\mathrm{U}k,i}\approx\sqrt{\beta} \left(\frac{1}{d_{\mathrm{U}k,i}}\right)^{\alpha/2}\mat{\tilde{H}}_{\mathrm{U}k,i}\label{eq:channelmatrix}
\end{align}
%where $\alpha$ is the path loss exponent, and 
%TODO $\beta$ represents a path loss at an implicated reference distance.
of dimension $N_R\times N_T$. The entries of the normalized channel matrix without path loss have an absolute value of $|(\mat{\tilde{H}}_i)_{kl}|=1$; the phase $\phase\{(\mat{\tilde{H}}_{\mathrm{U}K,i})\}_{mn}$ is independent and identically distributed (i.i.d.) on $[0,2\pi)$ for each $m\in\{1,\dots,N_R\}$ and $n\in\{1,\dots,N_T\}$. Moreover it is distributed identically on each possible UAV and user location while changing only moderate with small location changes.

%The exact phase can be known for the current position of the UAV only. It is impossible to predict it for different places.
Due to C-RAN, all UAVs fully collaborate via the ground BS through the high-capacity fronthaul links. Hence, the aggregate channel matrix from all users to the set of UAVs is given by
\begin{align}
\mat{H}_{i}=\begin{pmatrix}
\mat{H}_{\mathrm{U}1,i}^T&\dots&\mat{H}_{\mathrm{U}K,i}^T\end{pmatrix}^T.\label{eq:EquiChannelA}
\end{align}
Then, the noisy observation vector at the ground BS is given as
\begin{align}
\vect{v}=\sum_{i\in\mathcal{G}}\mat{H}_{i}\vect{u}_{i}+\vect{w},%_{\mathrm{U}}+\vect{w}_{\mathrm{C}},
\end{align}
where $\vect{u}_{i}$ is the transmit symbol vector of user $i$.
$\vect{w}$ represents the additive Gaussian noise at the UAV receivers combined with compression noise in fronthaul.
This is an independent Gaussian-distributed variables normalized in a way that $\vect{w}\sim\mathcal{CN}(\vect{0},\mat{I})$. We define the transmit signal covariance matrix as $\mat{Q}_i=\expectancy\{\vect{u}_{i}\vect{u}_{i}^H\}$.
%$\vect{w}_{\mathrm{U}}$ represents the additive Gaussian noise at the UAV receivers; $\vect{w}_{\mathrm{C}}$ denotes the compression noise in backhaul.
%Both are independent Gaussian-distributed variables, which are normalized in a way that $\vect{w}_{\mathrm{U}}+\vect{w}_{\mathrm{C}}\sim\mathcal{CN}(\vect{0},\mat{I})$. We define the transmit signal covariance matrix as $\mat{Q}_i=\expectancy\{\vect{u}_{i}\vect{u}_{i}^H\}$.

\subsection{Quadcopter UAV Model}
In control, systems are usually modelled as matrix differential equation with an additional input and a given initial condition
\begin{align}
\Sigma\begin{cases}
    \vect{\dot{s}}(t)=\mat{A}\vect{s}(t)+\mat{B}\vect{u}(t)\\
    \vect{s}(0)=\vect{s}_0,
    \end{cases}
\end{align}
where $\vect{s}(t)$ and $\vect{u}(t)$ refer to state and input vector at a given time $t$. The system model is described by $\mat{A}$ and $\mat{B}$ describing how each element of input and state vector impact the derivatives of the state vector elements.

In case of a quadcopter UAV, the system model uses the squared speeds of the four rotors as input values $\vect{u}_{k}(t)$. The state vector is described by location and orientation at a given point of time, as well as the derivatives of both. The location is described by the Cartesian coordinates $\vect{x}_{\mathrm{U}k}$; the orientation is described as $\vect{o}_{\mathrm{U}k}=\begin{pmatrix}-\theta_{k} & \phi_{k} & \psi_{k}\end{pmatrix}^T$, where $\theta_{k}$, $\phi_{k}$, and $\psi_{k}$ are the pitch, roll, and yaw  angles, respectively. Using a state vector of
$\vect{s}_{k}=\begin{pmatrix} \vect{x}_{\mathrm{U}k}^T & \vect{o}_{\mathrm{U}k}^T & \vect{\dot{x}}_{\mathrm{U}k}^T & \vect{\dot{o}}_{\mathrm{U}k}^T \end{pmatrix}^T$,
%$\vect{s}_{k}=\begin{pmatrix}x_{k} & y_{k} & z_{k} & \psi_{k} & \phi_{k} & \theta_{k} & \dot{x}_{k} & \dot{y}_{k} & \dot{z}_{k} & \dot{\psi}_{k} & \dot{\phi} & \dot{\theta}\end{pmatrix}^T$, 
a linearized model of a quadcopter can be derived as shown in \cite{7813499}. %, and is given in a similar form in \cite{Lunze16b}.
It can be phrased as
%a model of a quadcopter is given by \cite{Lunze16b} and can be phrased as follows.
\begin{align}
\tilde{\Sigma}_k:\begin{cases}
\vect{\dot{s}}_{k}(t)=\begin{pmatrix}
\zeros_{6\times 3}&\zeros_{6\times 1}&\zeros_{6\times 1}&\zeros_{6\times 1}&\eye_{6\times 6}\\
\zeros_{1\times 3}& g                &  0               & 0                &\zeros_{1\times 6}\\
\zeros_{1\times 3}& 0                &  g               & 0                &\zeros_{1\times 6}\\
\zeros_{4\times 3}&\zeros_{4\times 1}&\zeros_{4\times 1}&\zeros_{4\times 1}&\zeros_{4\times 6}\\
\end{pmatrix}\vect{{s}}_{k}(t)\\\hspace{2.5cm}+\begin{pmatrix}
\zeros_{8\times 4}\\\mat{\Theta}
\end{pmatrix}\vect{u}_{k}(t)+\begin{pmatrix}\zeros_{11\times 1}\\g\end{pmatrix}\\
%
%\vect{y}_{k}(t)=\begin{pmatrix}\eye_{4\times 4}&\zeros_{8\times 4}\end{pmatrix}\vect{{s}}_{k}(t)\\
%
\vect{{s}_{k}}(0)=\vect{{s}}_{k0}.
\end{cases}
\end{align}
%\begin{align}
%\Sigma:\begin{cases}
%%
%\vect{\dot{s}}_{k}(t)=\begin{pmatrix}
%\zeros_{6\times 4}&\zeros_{6\times 1}&\zeros_{6\times 1}&\eye_{6\times 6}\\
%\zeros_{1\times 4}& 0                & -g               &\zeros_{1\times 6}\\
%\zeros_{1\times 4}& g                &  0               &\zeros_{1\times 6}\\
%\zeros_{4\times 4}&\zeros_{4\times 1}&\zeros_{4\times 1}&\zeros_{4\times 6}\\
%\end{pmatrix}\vect{{s}}_{k}(t)\\\hspace{2.5cm}+\begin{pmatrix}
%\zeros_{8\times 4}\\\mat{\Theta}
%\end{pmatrix}\vect{u}_{k}(t)+\begin{pmatrix}\zeros_{8\times 1}\\g\\\zeros_{3\times 1}\end{pmatrix}\\
%%
%%\vect{y}_{k}(t)=\begin{pmatrix}\eye_{4\times 4}&\zeros_{8\times 4}\end{pmatrix}\vect{{s}}_{k}(t)\\
%%
%\vect{{s}_{k}}(0)=\vect{{s}}_{k0}
%%
%\end{cases}
%\end{align}
Here, $\mat{\Theta}$ is a device-specific full-rank-matrix; $g$ is the gravity constant. Even though it might not be that all states vector elements are given as output value, their values can be accessed through usage of an observer.

While controlling the input signal $\vect{u}_{k}(t)$, there is a chain of four integrators between the input signals and the horizontal parts of the UAV coordinates.

%The stabilization of the UAV on a constant height, as well as behavior upon distortions, and the energy distribution on the different rotors are separate control problems and ignored here.

When the input signal is chosen to be $\vect{u}_{k}(t)=\mat{\Theta}^{-1}\begin{pmatrix}u_{k,1}(t)&u_{k,2}(t)&u_z&0\end{pmatrix}^T$, the system in each horizontal direction $\gamma$ can be described as 
\begin{align}
\Sigma_{k,\gamma}:\begin{cases}
\vect{\dot{s}}_{\mathrm{U}k,\gamma}(t)\\\ \ \ \ \ =\begin{pmatrix}0&1&0&0\\0&0&g&0\\0&0&0&1\\0&0&0&0\end{pmatrix}\vect{s}_{\mathrm{U}k,\gamma}(t)+\begin{pmatrix}0\\0\\0\\1\end{pmatrix}u_{k,\gamma}(t)\\
%\begin{pmatrix}\dot{x}_{k}(t)\\\ddot{x}_{k}(t)\\\dot{\theta}_{k}(t)\\\ddot{\theta}_{k}(t)\end{pmatrix}&=\begin{pmatrix}0&1&0&0\\0&0&-g&0\\0&0&0&1\\0&0&0&0\end{pmatrix}\begin{pmatrix}{x}_{k}(t)\\\dot{x}_{k}(t)\\{\theta}_{k}(t)\\\dot{\theta}_{k}(t)\end{pmatrix}+\begin{pmatrix}0\\0\\0\\1\end{pmatrix}u_{xk}(t),\label{controlledSystemX}\\
%\begin{pmatrix}\dot{y}_{k}(t)\\\ddot{y}_{k}(t)\\\dot{\phi}_{k}(t)\\\ddot{\phi}_{k}(t)\end{pmatrix}&=\begin{pmatrix}0&1&0&0\\0&0&g&0\\0&0&0&1\\0&0&0&0\end{pmatrix}\begin{pmatrix}{y}_{k}(t)\\\dot{y}_{k}(t)\\{\phi}_{k}(t)\\\dot{\phi}_{k}(t)\end{pmatrix}+\begin{pmatrix}0\\0\\0\\1\end{pmatrix}u_{yk}(t)\label{controlledSystemY}.
\vect{s}_{\mathrm{U}k,\gamma}(0)=\vect{s}_{\mathrm{U}k,\gamma,0},
\end{cases}
\label{controlledSystem}
\end{align}
in which $\gamma\in\{1,2\}$ and
\begin{align}
    \vect{s}_{\mathrm{U}k,\gamma}(t)=\begin{pmatrix}\vect{x}_{\mathrm{U}k,\gamma}(t)\\\vect{\dot{x}}_{\mathrm{U}k,\gamma}(t)\\\vect{o}_{\mathrm{U}k,\gamma}(t)\\\vect{\dot{o}}_{\mathrm{U}k,\gamma}(t)\end{pmatrix}.\label{eq:statevectors}
\end{align}
%with output values $x$ and $y$. All other state values are determined using an observer.

\section{Achievable Spectral Efficiency (SE)}
A multiple access channel model is considered. The set of achievable rates of multiple users in cellular uplink channels is upper-bounded by \cite{Goldsmith:2005:WC:993515,Tse:2005:FWC:1111206}
\begin{align}
\sum_{i\in\mathcal{S}}R_{i}\leq&\frac{1}{2}\log\left|\mat{I}+\sum_{i\in\mathcal{S}}\mat{H}_{i}\mat{Q}_{i}\mat{H}_{i}^H\right|,\ \forall\mathcal{S}\subseteq\mathcal{G}, \label{fairnes2General}
\end{align}
in which the right-hand side is the capacity.
If the limitations of each upper-bound are divided equally among participating users, the minimum rate is upper-bounded by
%From this, the minimum rate can be obtained by the case of equal rate distribution among all users in $\mathcal{S}$ in each equation in above expression and is given by
\begin{align}
R_{\mathrm{min}}=&\min_{\mathcal{S}\subseteq\mathcal{G}}\frac{1}{2|\mathcal{S}|}\log\left|\mat{I}+\sum_{i\in\mathcal{S}}\mat{H}_{i}\mat{Q}_{i}\mat{H}_{i}^H\right|. \label{fairnes2Value}
\end{align}
Here, the subset actively bounding the optimization problem is given by $\mathcal{S}_{\mathrm{min}}=\arg_{\mathcal{S}\subseteq\mathcal{G}}R_{\mathrm{min}}$.
\section{Optimization Problem}
\label{sec:optimizationProblem}
The aim is to maximize this fairness rate. It depends on all UAV locations and the users' transmit signal covariance matrix.
%The data rate should be optimized with respect to users' position and transmit signal covariance matrix.
For a static placement, the optimization problem equals
\begin{align}
    \max_{\substack{\mat{Q}_1,\dots,\mat{Q}_{G}\\\vect{x}_1,\dots,\vect{x}_K}}&R_{\mathrm{min}}\left(\vect{x}_1,\dots,\vect{x}_K,\mat{Q}_1,\dots,\mat{Q}_{G}\right)\label{eq:optQ}\\
    \st\ \ &\trace\left\{\mat{Q}_i\right\}\leq P_{i,\mathrm{max}},\ \forall\ i\in\mathcal{G}\nonumber\\
    &\mat{Q}_i\succeq0,\ \forall\ i\in\mathcal{G}\nonumber\\
    & \vect{x}_{k,3}=h,\ \forall\ k\in\mathcal{K},\nonumber
\end{align}
in which $h$ is the altitude. Due to the change in impact of buildings, there is an optimal UAV altitude for achieving the best coverage as described in \cite{6863654}. Hence, $h$ is chosen to equal this value.%, and the channel matrices $\mat{H}_i$ depend on the coordinates as described in \eqref{eq:pathLoss}.

Since UAV motions will require some time, this is transformed into a dynamic optimization problem considering short intervals of time only. Hence, UAVs are only allowed to have moved a small distance from the previous location.
Over this time interval, the channels can be assumed to have negligible changes. Hence, the same covariance matrix is assumed to be optimal over this short time interval.

The trajectory and covariance matrices are optimized alternately, but can be updated in sampled time intervals only. Thus, the covariance matrices remain unchanged on each sampled flight interval. The location at the considered point of time of UAV $k$ is denoted as $\vect{\bar{x}}_k$; the optimal covariance matrix of user $i$ on this location is given as $\mat{\bar{Q}}_i$.
%Those are considered considered in the specific optimization problems.
%From this assumptions,
Using those, the optimization problem can be divided into two parts
\begin{align}
    \mat{\bar{Q}}_i = \argmax_{\mat{Q}_1,\dots,\mat{Q}_{G}}\,\ &R_{\mathrm{min}}\left(\vect{\bar{x}}_1,\dots,\vect{\bar{x}}_K,\mat{Q}_1,\dots,\mat{Q}_{G}\right)\label{eq:optQ}\\
    \st\ \ &\trace\left\{\mat{Q}_i\right\}\leq P_{i,\mathrm{max}},\ \forall\ i\in\mathcal{G}\nonumber\\
    &\mat{Q}_i\succeq0,\ \forall \ i\in\mathcal{G}\nonumber\\
    \vect{x}_k=\argmax_{\vect{{x}}_1,\dots,\vect{{x}}_K}\ \ &R_{\mathrm{min}}\left(\vect{{x}}_1,\dots,\vect{{x}}_K,\mat{\bar{Q}}_1,\dots,\mat{\bar{Q}}_{G}\right)\label{eq:optx}\\
    \st\ \ & \left|\left|\vect{x}_k-\vect{\bar{x}}_k\right|\right|\leq\varepsilon,\ \forall \ k\in\mathcal{K}\nonumber\\
     & \vect{x}_{k,3}=h,\ \forall \ k\in\mathcal{K},\nonumber
    %\nabla_k R_{\mathrm{min}} &= %\argmax_{x_k}
    %\frac{\partial R_{\mathrm{min}}\left(\vect{{x}}_1,\dots,\vect{{x}}_K,\mat{\bar{Q}}_1,\dots,\mat{\bar{Q}}_{|\mathcal{G}|}\right)}{\partial x_k},\label{eq:optgradx}
\end{align}
where $\varepsilon$ is an arbitrarily small number such that $R_{\mathrm{min}}$ can be approximated as an affine function in $\vect{x}_k$ for all $k$. Both problems can be solved in the given order in the interval of the sampling period. 

This principle can be illustrated as shown in Algorithm~\ref{alg:rate_traj_opt}, where at each sampling point both optimization problems are solved.

\begin{figure*}[!t]
	\centering
	\includegraphics{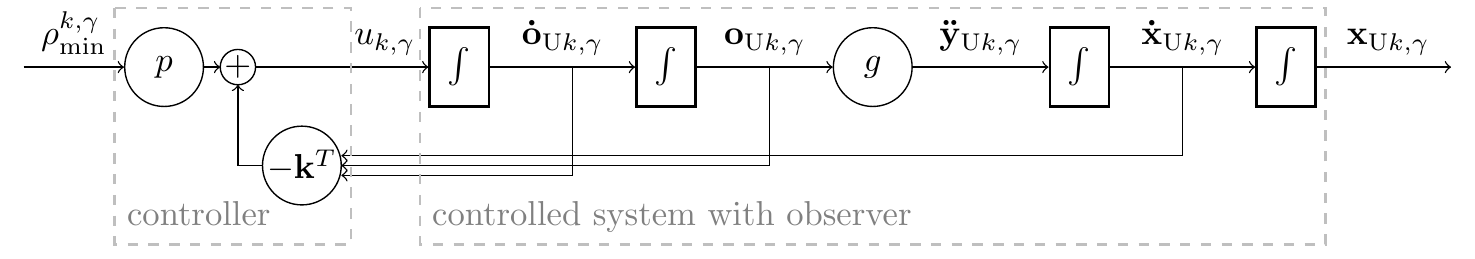}
	\caption{Control circuit for one Cartesian direction using P-controllers for navigating UAVs to a location with maximized minimum rate.}
	\label{fig:controlloops}
\end{figure*}

\begin{algorithm}[h]
\caption{Rate and trajectory optimization}
\begin{algorithmic}
\LOOP
\STATE $\mat{Q}_i \leftarrow$ result from \eqref{eq:optQ} $\forall i$
\STATE $\vect{x}_k \leftarrow$ result from \eqref{eq:optx} $\forall k$
%\STATE $\nabla_k R_{\mathrm{min}} \leftarrow$ result from \eqref{eq:optgradx} $\forall k$
\STATE control the UAV into the direction $\vect{x}_k$ seen from the current location of the UAV
\STATE wait for the sample time interval
\ENDLOOP
\end{algorithmic}
\label{alg:rate_traj_opt}
\end{algorithm}

It should be denoted that \eqref{eq:optQ} is a convex problem. However, \eqref{eq:optx} is non-convex, but since only a small interval is needed, an affine approximation can be used. This is done by calculating a data rate gradient,  %\eqref{eq:optgradx} describes a gradient, 
which is calculated in section~\ref{sec:dataRateGradient}. Algorithm~\ref{alg:rate_traj_opt} further bases on a UAV direction control, which is explained in section~\ref{sec:quadcopterUavControl}.

\section{Data Rate Gradient}
\label{sec:dataRateGradient}
For any arbitrary $\mathcal{S}\subseteq\mathcal{G}$, the derivative of the sum of rates from the set $\mathcal{S}$ given by \eqref{fairnes2General} w.r.t. the position coordinate of the $k$th UAV is given as
\begin{align}
&\rho_{\mathcal{S}}^{k,\gamma}=\frac{\partial}{\partial\vect{x}_{\mathrm{U}k,\gamma}}\sum_{i\in\mathcal{S}}R_{i}%=\frac{\partial}{\partial\vect{x}_{\mathrm{U}k,\gamma}}\frac{1}{2}\log\left|\mat{I}+\sum_{i\in\mathcal{S}}\mat{H}_{i}\mat{Q}_{i}\mat{H}_{i}^H\right|
\nonumber\\%,\ \forall\mathcal{S}\subseteq\mathcal{G}\\
&\hspace{0.5cm}=\frac{1}{2}\trace\left\{\left(\mat{I}+\sum_{i\in\mathcal{S}}\mat{H}_{i}\mat{Q}_{i}\mat{H}_{i}^H\right)^{-1}\sum_{i\in\mathcal{S}}\frac{\partial\mat{H}_{i}\mat{Q}_{i}\mat{H}_{i}^H}{\partial\vect{x}_{\mathrm{U}k,\gamma}}\right\},\label{fairnes2derivative}
\end{align}
which bases on \cite{IMM2012-03274}. Due to \eqref{eq:EquiChannelA}, the individual channel matrices $\mat{H}_{\mathrm{U}k,i}$ are implicated in this expression. Since the derivative of all of those is needed, through using \eqref{eq:channelmatrix}, this can be calculated as
\begin{align}
&\frac{\partial \mat{H}_{\mathrm{U}k,i}}{\partial\vect{x}_{\mathrm{U}k,\gamma}}=\sqrt{\beta}\frac{\partial}{\partial\vect{x}_{\mathrm{U}k,\gamma}}\frac{1}{d_{\mathrm{U}k,i}^{\alpha/2}} \mat{\tilde{H}}_{\mathrm{U}k,i}\nonumber\\
&= \frac{\sqrt{\beta}\alpha\left(\vect{x}_{i,\gamma}-\vect{x}_{\mathrm{U}k,\gamma}\right)}{2\left|\left|\vect{x}_{i}-\vect{x}_{\mathrm{U}k}\right|\right|^{\alpha/2+2}}\mat{\tilde{H}}_{\mathrm{U}k,i}+\frac{\sqrt{\beta}}{\left|\left|\vect{x}_{i}-\vect{x}_{\mathrm{U}k}\right|\right|^{\alpha/2}}\frac{\partial\mat{\tilde{H}}_{\mathrm{U}k,i}}{\partial\vect{x}_{\mathrm{U}k,\gamma}}\nonumber\\
&=\frac{\alpha}{2}\frac{\vect{x}_{i,\gamma}-\vect{x}_{\mathrm{U}k,\gamma}}{\left|\left|\vect{x}_{i}-\vect{x}_{\mathrm{U}k}\right|\right|^{2}}\mat{H}_{\mathrm{U}k,i}+\frac{\sqrt{\beta}}{\left|\left|\vect{x}_{i}-\vect{x}_{\mathrm{U}k}\right|\right|^{\alpha/2}}\frac{\partial\mat{\tilde{H}}_{\mathrm{U}k,i}}{\partial\vect{x}_{\mathrm{U}k,\gamma}}.
\end{align}
Please note that \eqref{eq:channelmatrix} has been used to simplify the first term of the result. Inserting this into \eqref{fairnes2derivative} leads to the derivative of the complete channel matrix of one user.

%\begin{align}
%\frac{\partial}{\partial\vect{x}_{\mathrm{U},\gamma}}\sum_{i\in\mathcal{S}}R_{i}&
%=\frac{1}{2}\trace\left\{\left(\mat{I}+\sum_{i\in\mathcal{S}}\mat{H}_{i}\mat{Q}_{i}\mat{H}_{i}^H\right)^{-1}\right.\nonumber\\&\left.\sum_{i\in\mathcal{S}}\left(\frac{\alpha\left(\vect{x}_{i,\gamma}-\vect{x}_{\mathrm{U},\gamma}\right)}{\left|\left|\vect{x}_{i}-\vect{x}_{\mathrm{U}}\right|\right|^{2}}\mat{H}_{i}\mat{Q}_{i}\mat{H}_{i}^H\right.\right.\nonumber\\&\left.\left.+\frac{\beta}{\left|\left|\vect{x}_{i}-\vect{x}_{\mathrm{U}}\right|\right|^{\alpha}}\frac{\partial\mat{\tilde{H}}_{i}\mat{Q}_{i}\mat{\tilde{H}}_{i}^H}{\partial\vect{x}_{\mathrm{U},\gamma}}\right)\right\}.
%\end{align}
Defining $\mat{\Phi}_{i}=\phase\{\mat{\tilde{H}}_{i}\}$ as the real-valued matrix of phases, the relation $\frac{\partial\mat{\tilde{H}}_{i}}{\partial\vect{x}_{\mathrm{U}k,\gamma}}=j\frac{\partial \mat{\Phi}_{i}}{\partial\vect{x}_{\mathrm{U}k,\gamma}}\circ \mat{\tilde{H}}_{i}$ holds, since, from \eqref{eq:channelmatrix}, the absolute value of each element of the normalized channel matrices is constant one. This means, the added multiplication factor $j\frac{\partial \mat{\Phi}_{i}}{\partial\vect{x}_{\mathrm{U}k,\gamma}}$ is imaginary solely. Since the same expression appears with the negative sign for the Hermitian matrix with identical absolute values and opposite sign, the sum of the traces of this two derivatives has little impact and vanishes for the SISO case. %In the MIMO case, we assume that the phase of the channel does not have significant impact on the UAV trajectories. Hence, we ignore this term for the trajectory.
In the MIMO case, we assume that the phase of the channel does have a random, not-significant impact only. Hence, we ignore this term for the trajectory.

Using the channel definition in \eqref{eq:channelmatrix} and the distance between UAV and users, the last part of \eqref{fairnes2derivative} can be reformulated as
%{\color{blue}\begin{align}
%&\frac{\partial}{\partial\vect{x}_{\mathrm{U}k,\gamma}}\sum_{i\in\mathcal{S}}R_{i}\nonumber\\
%&=\frac{1}{2}\trace\left\{\left(\mat{I}+\sum_{i\in\mathcal{S}}\mat{H}_{i}\mat{Q}_{i}\mat{H}_{i}^H\right)^{-1}\frac{\alpha}{2}\sum_{i\in\mathcal{S}}\frac{\vect{x}_{i,\gamma}-\vect{x}_{\mathrm{U}k,\gamma}}{\left|\left|\vect{x}_{i}-\vect{x}_{\mathrm{U}k}\right|\right|^{2}}\right.\nonumber\\&\left.\left(\begin{pmatrix}
%\zeros_{(k-1)N_R\times N_T}\\\mat{H}_{\mathrm{U}ki}\\\zeros_{(K-k)N_R\times N_T}
%\end{pmatrix}\mat{Q}_{i}\mat{H}_{i}^H+\mat{H}_{i}\mat{Q}_{i}\begin{pmatrix}
%\zeros_{(k-1)N_R\times N_T}\\\mat{H}_{\mathrm{U}ki}\\\zeros_{(K-k)N_R\times N_T}
%\end{pmatrix}^H\right)\right\}\nonumber\\&+\frac{1}{2}\trace\left\{\left(\mat{I}+\sum_{i\in\mathcal{S}}\mat{H}_{i}\mat{Q}_{i}\mat{H}_{i}^H\right)^{-1}\sum_{i\in\mathcal{S}}\mat{{H}}_{i}\frac{\partial\mat{Q}_{i}}{\partial\vect{x}_{\mathrm{U}k,\gamma}}\mat{{H}}_{i}^H\right\}
%\end{align}
%The remaining parts in the last term only depend on the derivative of $\mat{Q}_{i}$, a stochastic variables with constant trace $P_{i,\mathrm{max}}$. Thus, the second term is expected to have little impact on the final result only and vanishes for the SIMO case. Moreover, a UAV located at any specific position is not expected to know those derivatives in a usual case. Even if it did know, this factor could lead to unintended local maxima. When ignoring this term, and substituting back the channel matrices from \eqref{eq:channelmatrix}, the derivative can be approximated as} 
\begin{align}
&\sum_{i\in\mathcal{S}}\frac{\partial\mat{H}_{i}\mat{Q}_{i}\mat{H}_{i}^H}{\partial\vect{x}_{\mathrm{U}k,\gamma}}\approx\frac{\alpha}{2}\sum_{i\in\mathcal{S}}\frac{\vect{x}_{i,\gamma}-\vect{x}_{\mathrm{U}k,\gamma}}{\left|\left|\vect{x}_{i}-\vect{x}_{\mathrm{U}k}\right|\right|^{2}}\nonumber\\&\left(\begin{pmatrix}
\zeros_{(k-1)N_R\times N_T}\\\mat{H}_{\mathrm{U}ki}\\\zeros_{(K-k)N_R\times N_T}
\end{pmatrix}\mat{Q}_{i}\mat{H}_{i}^H+\mat{H}_{i}\mat{Q}_{i}\begin{pmatrix}
\zeros_{(k-1)N_R\times N_T}\\\mat{H}_{\mathrm{U}ki}\\\zeros_{(K-k)N_R\times N_T}
\end{pmatrix}^H\right).
\end{align}
%\begin{align}
%&\frac{\partial}{\partial\vect{x}_{\mathrm{U}k,\gamma}}\sum_{i\in\mathcal{S}}R_{i}
%\nonumber\\&=\frac{1}{2}\trace\left\{\left(\mat{I}+\sum_{i\in\mathcal{S}}\mat{H}_{i}\mat{Q}_{i}\mat{H}_{i}^H\right)^{-1}\frac{\alpha}{2}\sum_{i\in\mathcal{S}}\frac{\vect{x}_{i,\gamma}-\vect{x}_{\mathrm{U}k,\gamma}}{\left|\left|\vect{x}_{i}-\vect{x}_{\mathrm{U}k}\right|\right|^{2}}\right.\nonumber\\&\left.\left(\begin{pmatrix}
%\zeros_{(k-1)N_R\times N_T}\\\mat{H}_{\mathrm{U}ki}\\\zeros_{(K-k)N_R\times N_T}
%\end{pmatrix}\mat{Q}_{i}\mat{H}_{i}^H+\mat{H}_{i}\mat{Q}_{i}\begin{pmatrix}
%\zeros_{(k-1)N_R\times N_T}\\\mat{H}_{\mathrm{U}ki}\\\zeros_{(K-k)N_R\times N_T}
%\end{pmatrix}^H\right)\right\}\label{eq:derivativeApprox2}
%\end{align}
Note that the channel $\mat{H}_{i}$ depends on the location of the UAV at the specific moment of time.
Because the transmit signal covariance matrix $\mat{Q}_{i}$ is constant within the sampling interval, it does not appear in above expression.
%The derivative of the transmit signal covariance matrix $\mat{Q}_{i}$ does not appear in above expression, since this is set constant during a considered sampling interval.

Using \eqref{fairnes2Value}, the gradient of the achievable minimum rate \eqref{fairnes2derivative} can be calculated as 
\begin{align}
\frac{\partial R_{\mathrm{min}}}{\partial\vect{x}_{\mathrm{U}k,\gamma}}%&=\frac{1}{|\mathcal{S}_{\mathrm{min}}|}\frac{\partial}{\partial\vect{x}_{\mathrm{U}k,\gamma}}\sum_{i\in\mathcal{S_{\mathrm{min}}}}R_{i}\nonumber\\
&=\frac{\rho_{\mathcal{S}_{\mathrm{min}}}^{k,\gamma}}{|\mathcal{S}_{\mathrm{min}}|}=:\rho_{\mathrm{min}}^{k,\gamma}.
\end{align}
Using this, the gradient can be phrased as a three-dimensional vector $\nabla_{k} R_{\mathrm{min}}$ containing the elements
\begin{align}
\left(\nabla_{k} R_{\mathrm{min}}\right)_\gamma
=\rho_{\mathrm{min}}^{k,\gamma}
%=\frac{\partial R_{\mathrm{min}}}{\partial\vect{x}_{\mathrm{U}{k},\gamma}}
.\label{eq:gradApproxMinRate}
\end{align}

\begin{figure*}[t]
    \centering
    \includegraphics{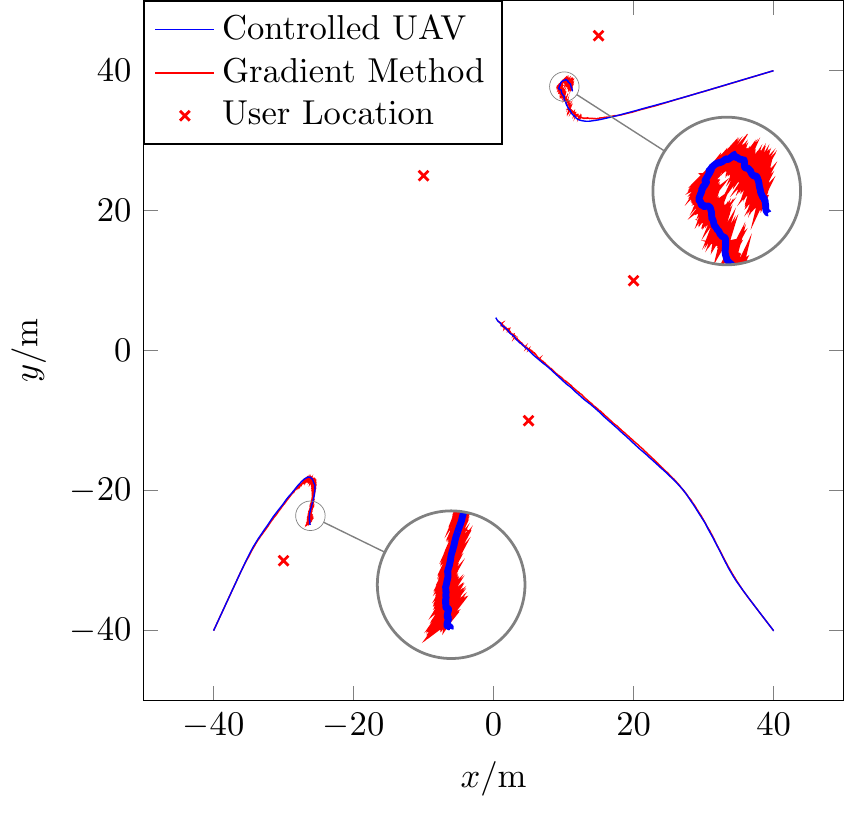}
    \hspace{0.4cm}
    \includegraphics{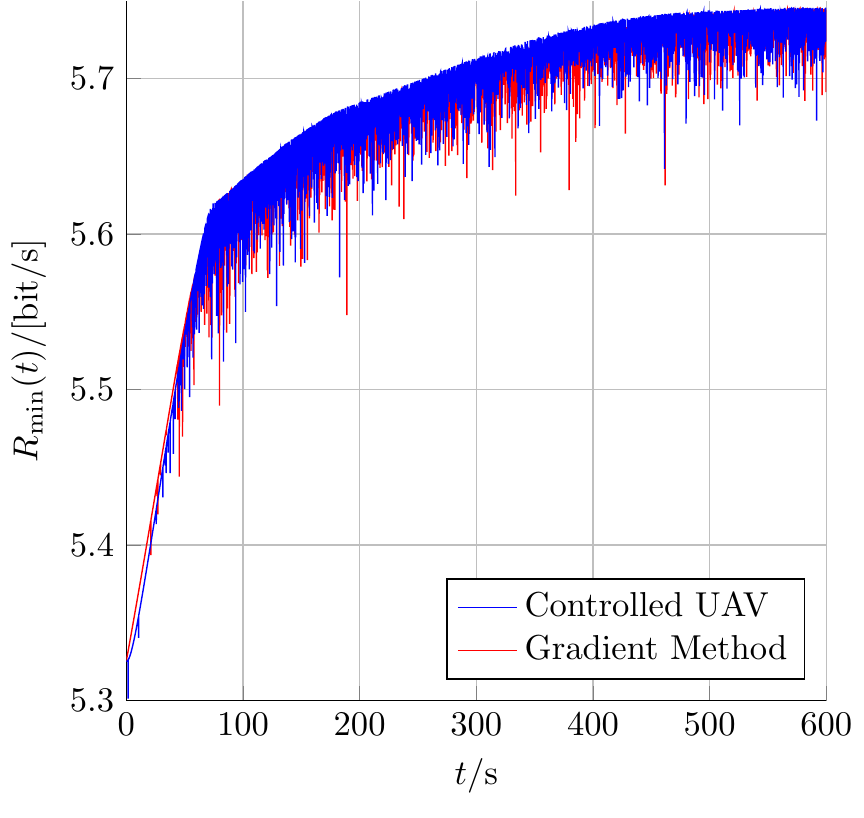}
    \caption{Trajectory and minimum rate for controlled UAVs ideal gradient method. The UAVs start in the corners and move then with decreasing velocity into the direction of the users. Receiving UAVs have eight antennas, transmitting users one. UAV altitude is \SI{50}{\metre}.}
    \label{fig:numerical_results}
\end{figure*}

\section{Quadcopter UAV Control}
\label{sec:quadcopterUavControl}
The UAV locations should remain unchanged when a rate maximum is obtained. Thus, the static end value is supposed to fulfill the expression $\nabla_{k}R_{\mathrm{min}}=0\Rightarrow \vect{\dot{x}}_{\mathrm{U}k}=\zeros$. While the rate gradient does not equate to zero, the UAV should be navigated into the direction of the gradient.

In order to fulfill this, the input values $u_{k,\gamma}(t)$ should be controlled in a way such that the static end values of $\vect{\dot{x}}_{\mathrm{U}k,\gamma}$ match for constant rate derivatives their corresponding entries of $\nabla_{k} R_{\mathrm{min}}$ for all $\gamma\in\{1,2\}$.
By using a P-controller, the input signal is chosen to be
\begin{align}
u_{k,\gamma}=& p
\left(\nabla_{k} R_{\mathrm{min}}\right)_\gamma
-\vect{k}^T\begin{pmatrix}\vect{\dot{x}}_{\mathrm{U}k,\gamma}(t)\\\vect{o}_{\mathrm{U}k,\gamma}(t)\\\vect{\dot{o}}_{\mathrm{U}k,\gamma}(t)\end{pmatrix},\forall\gamma\in\{1,2\},\label{Controller}
\end{align}
%The sign change is due to the condition for negative feedback in the complete control loop. 
in which $\vect{k}=\begin{pmatrix}k_1&k_2&k_3\end{pmatrix}^T$ is the controller gain; $p$ refers to the prefilter. In the vector used for feedback, $\vect{x}_{\mathrm{U}k,\gamma}$ does not occur. This is because the velocity should be controlled to zero, not the location itself. An individual, combined control circuit is shown in \figurename~\ref{fig:controlloops} on top of the previous page.

In a final system, the controller matrices depends on aspects such as altitude control and disturbance cancellation, and thus, i.e., they can not be chosen completely freely. In the numeric results given here, they are determined using a linear-quadratic regulator (LQR).
	
The overall system can be described by combining \eqref{controlledSystem} and \eqref{Controller}. Then, the system parameters can be derived as
\begin{align}
\mat{A}=\begin{pmatrix}0&1&0&0\\0&0&g&0\\0&0&0&1\\0&-k_{1}&-k_{2}&-k_{3}\end{pmatrix}&,\ \ \ \vect{b}=\begin{pmatrix}0\\0\\0\\p\end{pmatrix},
\end{align}
the state vectors are given in \eqref{eq:statevectors}.

%\begin{align}
%\vect{s}_{xk}=&\begin{pmatrix}{x}_{k}(t)&\dot{x}_{k}(t)&{\theta}_{k}(t)&\dot{\theta}_{k}(t)\end{pmatrix}^T,\\ \vect{s}_{yk}=&\begin{pmatrix}{y}_{k}(t)&\dot{y}_{k}(t)&{\phi}_{k}(t)&\dot{\phi}_{k}(t)\end{pmatrix}^T.
%\end{align}
From this, the trajectory of each controlled UAV can be described using the state equation
\begin{align}
\begin{pmatrix}\vect{\dot{s}}_{\mathrm{U}k,1}\\\vect{\dot{s}}_{\mathrm{U}k,2}\end{pmatrix}
=&\left(\eye_{2}\otimes\mat{A}\right)
\begin{pmatrix}\vect{s}_{\mathrm{U}k,1}\\\vect{s}_{\mathrm{U}k,2}\end{pmatrix}\nonumber\\&
+\left(\begin{pmatrix}
1 & 0 & 0\\
0 & 1 & 0\\
\end{pmatrix}\otimes\vect{b}\right)\nabla_{k} R_{\mathrm{min}}.
\end{align}
Since the UAV altitude is supposed to remain constant, the vertical dimension of the gradient is multiplied with zero only in above expression.

Each individual UAV $k\in\left\{1,\dots K\right\}$ can be described by this equation. Combined, this describes the full system of UAVs, which are used for rate maximization on locations determined by control methods.
\section{Numerical Results}
\label{sec:numericalResults}
For the numerical results here, the controllers are determined using a LQR. This way, their values are set to $\vect{k}^T=\begin{pmatrix}
\SI{0.5477}{\metre\second}  & \SI{23.9683}{\second^2}  &  \SI{6.9308}{\second}
\end{pmatrix}$. The initial states are set to $\vect{s}_{\mathrm{U}k,\gamma,0}=\begin{pmatrix}\vect{x}_{\mathrm{U}k,\gamma,0}& 0& 0& 0\end{pmatrix}^T$, where $\vect{x}_{\mathrm{U}k,\gamma,0}=\pm\SI{40}{\metre}$ as shown in the corners of \figurename~\ref{fig:numerical_results}. The overall control system is discretized, such that the rate gradients only need to be known in short time intervals.

For comparison, additional simulations are done with a more simple gradient method. There, the velocities are set directly proportional to the rate gradient. The integration of those over time equals the location of the UAV. This system does not contain any feedback paths, and there is no controller to be designed. This is a mathematical idealized model for the trajectory. Thereby, the UAV trajectory can perform fast direction and velocity changes. This comparison system is without practical relevance for real-time applications, but might be relevant in scenarios where full trajectories are planned beforehand. The trajectories and data rates over time are compared for both methods in \figurename~\ref{fig:numerical_results}.

%This results show that the UAVs move over time to locations, where higher data rates are achievable.
In both simulations, the different UAVs are placed on locations apart from each other leading to having a higher coverage area. The control system adapts slower to gradient changes than the gradient method. This comes with a rather slowly growth of the data rate at the beginning, but leads to a more smooth trajectory with less noise near the optimal points.

\section{Conclusion and Future Work}
The presented algorithm optimizes the UAV placements and trajectories online while serving dynamically located users with data rates. For the static case, the UAVs reach a local optimal set of positions and remain there until the user locations change. This has been achieved throug applying control methods on the minimum rate maximization problem. This methods only require information about channels and locations for the given moment of time. Additional Information are not required.
%The numerical results show that it is possible to use control methods to determine the best locations for minimum rate maximization and relocate UAVs there. This does not require knowledge about CSI on any locations on the map other the ones where the UAVs and users are actually placed.
The approach presented here is directly suitable for real-time applications, as it can adapt to changes of user locations and channel coefficients instantaneously. There is no exhaustive search in use. Nevertheless, the UAVs find a good location to be placed for a permanent rate transfer. From the numerical results, this location is even more stable with the control model than with the mathematical gradient method.

This concept works for homogeneous areas, such as rural areas or concert places. In future work, this will be extended by considering heterogeneous building maps, where the channel amplitude is variable due to non-constant LoS connections. A combination of statistical approaches basing on user densities and this type of real-time adaption might be useful to achieve further enhancements in dense urban environments.

% conference papers do not normally have an appendix

% use section* for acknowledgment
%\section*{Acknowledgment}

%The authors would like to thank...

% trigger a \newpage just before the given reference
% number - used to balance the columns on the last page
% adjust value as needed - may need to be readjusted if
% the document is modified later
%\IEEEtriggeratref{8}
% The "triggered" command can be changed if desired:
%\IEEEtriggercmd{\enlargethispage{-5in}}

% references section

% can use a bibliography generated by BibTeX as a .bbl file
% BibTeX documentation can be easily obtained at:
% http://mirror.ctan.org/biblio/bibtex/contrib/doc/
% The IEEEtran BibTeX style support page is at:
% http://www.michaelshell.org/tex/ieeetran/bibtex/
\bibliographystyle{IEEEtran}
% argument is your BibTeX string definitions and bibliography database(s)
\bibliography{bibliography}
%
% <OR> manually copy in the resultant .bbl file
% set second argument of \begin to the number of references
% (used to reserve space for the reference number labels box)
%\begin{thebibliography}{1}
%\end{thebibliography}

% that's all folks
\end{document}

%% file: header.tex
\usepackage{siunitx}
\usepackage{amsmath}
\usepackage{amsfonts}
\usepackage{dsfont}

\DeclareMathOperator*{\argmax}{\mathrm{arg\,max}}

\newcommand{\mat}[1]{\mathbf{#1}}
\newcommand{\vect}[1]{\mathbf{#1}}
\newcommand{\trace}{\mathrm{trace}}
\newcommand{\phase}{\mathrm{phase}}
\newcommand{\expectancy}{{\mathds{E}}}
\newcommand{\Gaussian}{\mathcal{N}}

\newcommand{\st}{\mathrm{subject\ to}}
\newcommand{\eye}{\mat{I}}
\newcommand{\zeros}{\mat{0}}
\def\mobile{\leavevmode\hbox to7bp{\kern1bp \lower1bp\vbox to12bp{}%
		\pdfliteral{q 0 g 0 G 1 j 2 w 0 0 5 10 re B
			1 g 1 G 1 w .3 1.8 4.4 7 re B 
			1.5 w 2.5 .2 0 .1 re B .3 w 1.7 10 1.6 0 re B Q}%
		\hss}}

%% file: bare_conf.bbl
% Generated by IEEEtran.bst, version: 1.14 (2015/08/26)
\begin{thebibliography}{10}
\providecommand{\url}[1]{#1}
\csname url@samestyle\endcsname
\providecommand{\newblock}{\relax}
\providecommand{\bibinfo}[2]{#2}
\providecommand{\BIBentrySTDinterwordspacing}{\spaceskip=0pt\relax}
\providecommand{\BIBentryALTinterwordstretchfactor}{4}
\providecommand{\BIBentryALTinterwordspacing}{\spaceskip=\fontdimen2\font plus
\BIBentryALTinterwordstretchfactor\fontdimen3\font minus
  \fontdimen4\font\relax}
\providecommand{\BIBforeignlanguage}[2]{{%
\expandafter\ifx\csname l@#1\endcsname\relax
\typeout{** WARNING: IEEEtran.bst: No hyphenation pattern has been}%
\typeout{** loaded for the language `#1'. Using the pattern for}%
\typeout{** the default language instead.}%
\else
\language=\csname l@#1\endcsname
\fi
#2}}
\providecommand{\BIBdecl}{\relax}
\BIBdecl

\bibitem{8316776}
S.~Sekander, H.~Tabassum, and E.~Hossain, ``Multi-tier drone architecture for
  {5G/B5G} cellular networks: Challenges, trends, and prospects,'' \emph{IEEE
  Communications Magazine}, vol.~56, no.~3, pp. 96--103, March 2018.

\bibitem{DBLP:journals/corr/abs-1803-00680}
\BIBentryALTinterwordspacing
M.~Mozaffari, W.~Saad, M.~Bennis, Y.~Nam, and M.~Debbah, ``A tutorial on {UAVs}
  for wireless networks: Applications, challenges, and open problems,''
  \emph{CoRR}, vol. abs/1803.00680, 2018. [Online]. Available:
  \url{http://arxiv.org/abs/1803.00680}
\BIBentrySTDinterwordspacing

\bibitem{DBLP:journals/corr/abs-1711-07668}
\BIBentryALTinterwordspacing
P.~Chandhar and E.~G. Larsson, ``Massive {MIMO} for drone communications:
  Applications, case studies and future directions,'' \emph{CoRR}, vol.
  abs/1711.07668, 2017. [Online]. Available:
  \url{http://arxiv.org/abs/1711.07668}
\BIBentrySTDinterwordspacing

\bibitem{8048502}
A.~Al-Hourani and K.~Gomez, ``Modeling cellular-to-{UAV} path-loss for suburban
  environments,'' \emph{IEEE Wireless Communications Letters}, vol.~7, no.~1,
  pp. 82--85, Feb 2018.

\bibitem{8376956}
E.~Koyuncu, R.~Khodabakhsh, N.~Surya, and H.~Seferoglu, ``Deployment and
  trajectory optimization for {UAVs}: A quantization theory approach,'' in
  \emph{2018 IEEE Wireless Communications and Networking Conference (WCNC)},
  April 2018, pp. 1--6.

\bibitem{8292533}
J.~Kakar and V.~Marojevic, ``Waveform and spectrum management for unmanned
  aerial systems beyond 2025,'' in \emph{2017 IEEE 28th Annual International
  Symposium on Personal, Indoor, and Mobile Radio Communications (PIMRC)}, Oct
  2017, pp. 1--5.

\bibitem{DBLP:journals/corr/abs-1805-07822}
\BIBentryALTinterwordspacing
J.~Kakar, A.~Chaaban, V.~Marojevic, and A.~Sezgin, ``{UAV}-aided multi-way
  communications,'' \emph{CoRR}, vol. abs/1805.07822, 2018. [Online].
  Available: \url{http://arxiv.org/abs/1805.07822}
\BIBentrySTDinterwordspacing

\bibitem{5461502}
A.~I. Alshbatat and L.~Dong, ``Cross layer design for mobile ad-hoc unmanned
  aerial vehicle communication networks,'' in \emph{2010 International
  Conference on Networking, Sensing and Control (ICNSC)}, April 2010, pp.
  331--336.

\bibitem{8255739}
C.~Lin, D.~He, N.~Kumar, K.~R. Choo, A.~Vinel, and X.~Huang, ``Security and
  privacy for the internet of drones: Challenges and solutions,'' \emph{IEEE
  Communications Magazine}, vol.~56, no.~1, pp. 64--69, Jan 2018.

\bibitem{7486987}
M.~Mozaffari, W.~Saad, M.~Bennis, and M.~Debbah, ``Efficient deployment of
  multiple unmanned aerial vehicles for optimal wireless coverage,'' \emph{IEEE
  Communications Letters}, vol.~20, no.~8, pp. 1647--1650, Aug 2016.

\bibitem{DBLP:journals/corr/MozaffariSBD15a}
\BIBentryALTinterwordspacing
------, ``Drone small cells in the clouds: Design, deployment and performance
  analysis,'' \emph{CoRR}, vol. abs/1509.01655, 2015. [Online]. Available:
  \url{http://arxiv.org/abs/1509.01655}
\BIBentrySTDinterwordspacing

\bibitem{6863654}
A.~Al-Hourani, S.~Kandeepan, and S.~Lardner, ``Optimal lap altitude for maximum
  coverage,'' \emph{IEEE Wireless Communications Letters}, vol.~3, no.~6, pp.
  569--572, Dec 2014.

\bibitem{7762053}
J.~Lyu, Y.~Zeng, R.~Zhang, and T.~J. Lim, ``Placement optimization of
  {UAV}-mounted mobile base stations,'' \emph{IEEE Communications Letters},
  vol.~21, no.~3, pp. 604--607, March 2017.

\bibitem{8011325}
Z.~Becvar, M.~Vondra, P.~Mach, J.~Plachy, and D.~Gesbert, ``Performance of
  mobile networks with {UAVs}: Can flying base stations substitute ultra-dense
  small cells?'' in \emph{European Wireless 2017; 23th European Wireless
  Conference}, May 2017, pp. 1--7.

\bibitem{8385464}
O.~Esrafilian and D.~Gesbert, ``Simultaneous user association and placement in
  multi-{UAV} enabled wireless networks,'' in \emph{WSA 2018; 22nd
  International ITG Workshop on Smart Antennas}, March 2018, pp. 1--5.

\bibitem{1563300}
A.~Harris, J.~J. Sluss, H.~H. Refai, and P.~G. LoPresti, ``Alignment and
  tracking of a free-space optical communications link to a {UAV},'' in
  \emph{24th Digital Avionics Systems Conference}, vol.~1, Oct 2005, pp.
  1.C.2--1.1.

\bibitem{7813499}
P.~Wang, Z.~Man, Z.~Cao, J.~Zheng, and Y.~Zhao, ``Dynamics modelling and linear
  control of quadcopter,'' in \emph{2016 International Conference on Advanced
  Mechatronic Systems (ICAMechS)}, Nov 2016, pp. 498--503.

\bibitem{Goldsmith:2005:WC:993515}
A.~Goldsmith, \emph{Wireless Communications}.\hskip 1em plus 0.5em minus
  0.4em\relax New York, NY, USA: Cambridge University Press, 2005.

\bibitem{Tse:2005:FWC:1111206}
D.~Tse and P.~Viswanath, \emph{Fundamentals of Wireless Communication}.\hskip
  1em plus 0.5em minus 0.4em\relax New York, NY, USA: Cambridge University
  Press, 2005.

\bibitem{IMM2012-03274}
\BIBentryALTinterwordspacing
K.~B. Petersen and M.~S. Pedersen, ``The matrix cookbook,'' nov 2012, version
  20121115. [Online]. Available: \url{http://www2.imm.dtu.dk/pubdb/p.php?3274}
\BIBentrySTDinterwordspacing

\end{thebibliography}
